\documentclass[useAMS,usenatbib]{mn2e}
\usepackage{graphics}
\usepackage{epsfig}
\usepackage{natbibmnfix}
\usepackage{natbib}
\newcommand{\hMpc}{\mbox{$h^{-1}$ Mpc} }

\title[Voids and Void Galaxies]{Void Statistics and Void Galaxies in the 2dFGRS}
\author[]{Alexander M. von Benda-Beckmann$^{1}$, Volker M\"uller$^{1}$
\\
$^{1}$Astrophysical Institute Potsdam, An der Sternwarte 16, Germany}

\begin{document}

\date{re-submitted Version 2007 october}

\pagerange{\pageref{firstpage}--\pageref{lastpage}} \pubyear{2006}

\maketitle

\label{firstpage}

\begin{abstract}
For the 2dFGRS we study the properties of voids and of fainter galaxies within 
voids that are defined by brighter galaxies. Our results are compared with 
simulated galaxy catalogues from the Millenium simulation coupled with a 
semianalytical galaxy formation recipe. We derive the void size distribution 
and discuss its dependence on the faint magnitude limit of the galaxies
defining the voids. While voids among faint galaxies are typically smaller
than those among bright galaxies, the ratio of the void sizes 
to the mean galaxy separation reaches larger values. This is well
reproduced in the mock galaxy samples studied. We provide analytic fitting 
functions for the void size distribution. Furthermore, we study the galaxy 
population inside voids defined by objects with $B_J -5\log{h}< -20$ and diameter 
larger than 10 \hMpc. We find a clear bimodality of the void galaxies similar 
to the average comparison sample. We confirm the enhanced abundance of
galaxies in the blue cloud and a depression of the number of red sequence 
galaxies. There is an indication of a slight blue shift of the blue cloud. 
Furthermore, we find that galaxies in void centers have higher specific star 
formation rates as measured by the $\eta$ parameter. We determine the radial 
distribution of the ratio of early and late type galaxies through the voids. 
We find and discuss some differences between observations and the Millenium 
catalogues.
\end{abstract}


\section{Introduction}

The large-scale galaxy distribution is highly inhomogeneous. We observe
groups, clusters and superclusters of galaxies and large voids. During 
last decades, much attention was paid on the analysis of bound structures
as groups and clusters. Recently, new superclusters catalogues were 
constructed from the 2dFGRS and compared with large cosmological 
simulations \citep{Einasto07a, Einasto07b}. In a complement, 
there are large regions in the universe without bright galaxies, so 
called cosmic voids. Early on very large voids over 50 \hMpc diameter were 
found by \citet{Gregory78} and \citet{Kirshner81}. More common are voids with 
diameters of about 10 \hMpc that fill most of cosmic space. 
The explanation of such structures is not obvious. According to the 
standard paradigm of cosmological structure formation, negative potential 
wells from primordial inhomogeneities attract all matter in bound structures. 
In the same way, positive potential perturbations expel matter, but observed
voids are too large for completely emptying. Therefore, in addition 
to the dilution of matter, the galaxy formation probability should be 
suppressed in underdense regions, 
cp. e.g. \citet{Lee98, Madsen98}. Recently, \citet{Sheth04} and 
\citet{Furlanetto06} applied these ideas within the excursion set
formalism of gravitational instability. These analytical theories derived 
void size distributions that are peaked typically at diameters below 
10 \hMpc which seem to be smaller than observed void sizes,
cp. \citet{Mueller00}, and void sizes in CDM-simulations 
coupled with semianalytical galaxy formation models, \citet{Benson03}. 

Voids were routinely identified in all wide-field redshift surveys as the 
CfA \citep{deLapparent86, Vogeley94}, the SSRS2 \citep{Elad97}, 
the LCRS \citep{Mueller00, Arbabi02}, the IRAS-survey \citep{Elad00}, 
the 2dFGRS \citep{Hoyle04, Croton04, Patiri06a}, 
the SDSS \citep{Rojas04, Rojas05, Patiri06b}, 
and the DEEP2 survey with an analysis 
of voids up to redshift $z \approx 1$ \citep{Conroy05}. However, 
many void searches are only devoted to the identification of large voids, 
other void finders depend on special procedures as firstly
identifying wall galaxies by an overdensity criterion and then looking for
voids bounded by wall galaxies \citep{Elad97, Hoyle04}. Furthermore, the 
void search
depends on the galaxy sample used for defining voids, in particular on 
the limiting magnitude of the galaxy sample. In an influential paper, 
\citet{Peebles01} derived from nearest neighbor statistics that
galaxies of different brightness respect the same voids. He claimed 
that this contradicts the standard CDM scenario of galaxy and 
structure formation that seem to predict a hierarchy of galactic structures
with smaller structure for fainter objects sitting in less massive dark matter
halos, i.e also smaller voids for fainter objects. 
In a follow up theoretical study, \citet{Mathis02} 
showed from high-resolution simulation that voids defined by bright
galaxies are also underdense in faint galaxies, i.e. that bright and faint
galaxies respect similar voids. We want to take up this question 
since our earlier studies of voids in LCRS and in LCDM-simulations
\citep{Mueller00, Arbabi02} showed a dependence of the void size distribution 
on the brightness limit of the galaxies under study and a characteristic 
void size scaling relation. \citet{Benson03} confirmed this scaling relation 
in simulated galaxy distributions, but the quantitative parameters were 
different. They suspected differences in the void search algorithms as reason, 
but we suspect that the effective 2-dimensional nature of the LCRS is 
the most likely cause. But their demand for using the same void search
algorithm both in data and simulations seems a prerequisite for trustworth 
results. More recently, Colberg et al. (2007 in preparation) 
compared different void search
algorithms and found that most proposed algorithm find comparable locations 
and sizes of large voids. This is very likely not the case for the 
large number of small voids that fill a significant part of space. Therefore 
we shall present in this study an comparable analysis of voids both in
simulations and in the data that explores the detailed void size distribution 
in dependence on the faint brightness limit of the galaxies defining the voids.

We shall use for our study the 2dFGRS \citep{Cole05} thereby coming back
to the property of the self-similarity of the void statistics. It tells
that the void size distribution depends on the mean galaxy separation, 
in such a way that brighter galaxies define larger voids than fainter ones. 
Even if voids in the 2dFGRS were previously analyzed \citep{Hoyle04, 
Patiri06a}, this concerned mainly large voids and not the detailed 
void statistics proposed by us previously. \citet{Croton04} provided a 
detailed study of the void probability distribution for the 2dFGRS which is 
related to the void size distribution but it provides an different
statistics. Essentially the void probability distribution is a weighted 
sum over the void size distribution \citep{Otto86}. 

The 2dFGRS is a densely sampled survey with a compact survey geometry. 
This is of advantage for the question of the dependence of the void 
sizes on the galaxy magnitudes defining voids. We shall derive
phenomenological fits to the void size distribution that will be 
compared with simulation results. Furthermore, we shall take up the 
question of the faint galaxies within voids. Thereby we cut both questions, 
the matter content inside large underdense regions in the universe, and the 
change of galaxy properties. The color distribution of galaxies in the
2dFGRS employs SuperCOSMOS data \citep{Hambly01} for the $R$-band. We will 
find a clear bimodality in the void galaxies so far only studied in detail 
for the SDSS \citep{Rojas05, Patiri06b} but not yet for the 2dFGRS.

We shall evaluate our void analysis with model galaxy samples constructed from 
the Millenium simulation of \citet{Springel05} and from semianalytical 
galaxy formation theory applied to the numerical merger 
trees \citep{Croton06}. We analyzed specific galaxy 
properties within voids and found results that can be qualitatively 
described by the model samples. A quantitative comparison 
of the galaxy color distribution and the star formation 
efficiency hints at certain differences between observations and 
simulations. 
Tentatively we connect it with specific environmental properties of 
galaxy formation in underdense regions. In particular, major mergers, 
galaxy harassment, tidal and ram-pressure stripping will not be 
as effective there as in more dense regions of the universe \citep{Avila05, 
Maulbetsch06}.

The outline of the paper is as follows: First we describe the galaxy
extraction from the 2dFGRS, and in Section 3 we provide some details of 
the galaxy mock data. In Section 4 we describe our void search algorithm 
and in Section 5 we provide our results. Section 6 is devoted to 
a discussion and in the final Section we draw our conclusions.

\section{Selection of 2dFGRS data}

We analyze the 2dF Galaxy Redshift Survey (2dFGRS, \citet{Colless01}) with
about 222,000 galaxies covering a sky area of 1500 deg$^2$. For the void
search and statistics, it has the advantage of covering a sufficient large
area with an average completeness of $90\%$ in redshift measurements,
\citet{Colless01}. We shall discuss the method for treating the varying
completeness below. Photometric data in $b_{J}$ stem from the APM photographic
plates \citep{Maddox90} which represents the basis of the fiber placement for
redshift measurements. $b_J$ and $r$-band data are available through the
SuperCOSMOS catalog (\cite{Hambly01}).

\begin{figure} 
 \epsfig{file=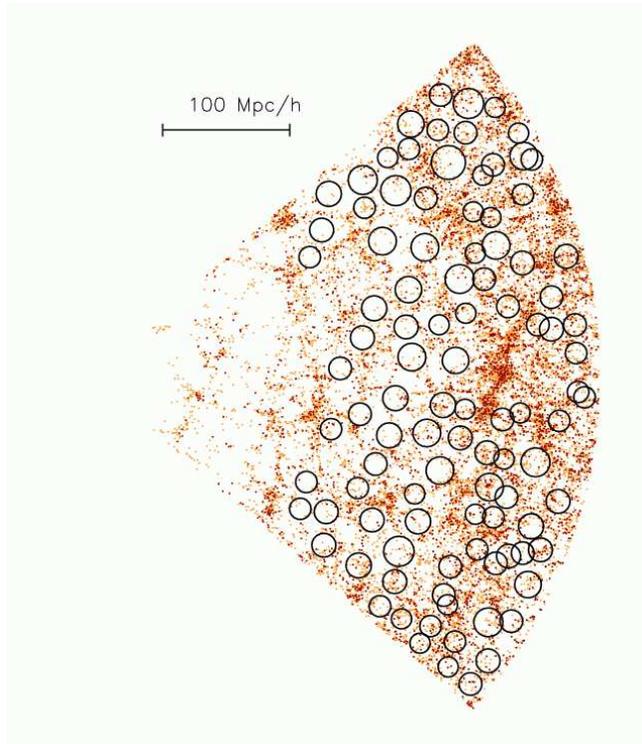,width=8.5cm}
 \vspace{0.3cm}
\caption{Projected distribution of the 100 largest voids in 2dFGRS S3
  sample for $B_{J} < -20$. Black (red in the electronic version of the paper)
  symbols are galaxies brighter than the magnitude threshold, and grey
  (orange) symbols fainter galaxies that partly fill the voids. Note that same
  bright galaxies appear inside the voids due to projection effects.}
\label{voids2df}
\end{figure}

\begin{figure}
  \epsfig{file=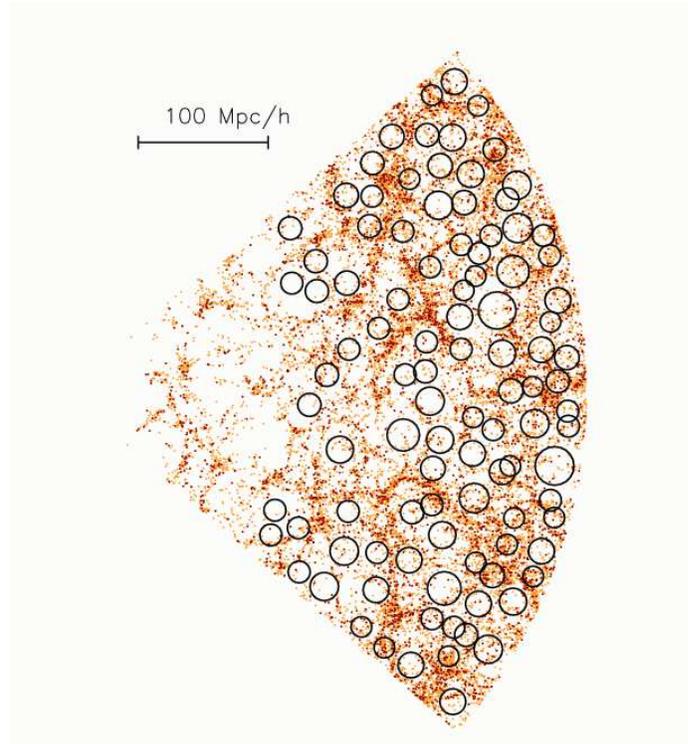,width=9cm} 
  \vspace{0.4cm}
\caption{Projected distribution of 100 largest voids in a Millenium mock
  catalog MW3. The color coding is identical to Fig.\ref{voids2df}.}
\label{voidsmill}
\end{figure}

The 2dFGRS has been used for identifying large voids by \citet{Hoyle04} and by
\citet{Patiri06a}, and for an analysis of the void probability function by
\cite{Croton04}. Our aim is to touch both points. We extract a large catalogue
of voids, both small and large ones. This catalogue can be used for
statistical studies of the void size distribution, and for identifying
subpopulations within voids. We do not claim that all our voids are
statistically significant. Especially the position of the large number of
small voids are influenced by random galaxy positions. But a detailed look at
the large voids in Figs. \ref{voids2df} and \ref{voidsmill} verifies that both
in data and simulations we catch as voids the same regions as the eye selects
in the galaxy distribution (note the projection effects that influences a part
of voids in this Figure).
  
We extracted volume limited samples from the 2dFGRS in using the $b_J$
magnitudes since it is the color defining the spectroscopic survey.  Apparent
magnitudes were converted to absolute magnitudes by using the extinction
corrections as given in the 2dFGRS catalog for the APM $b_{J}$ magnitudes. The
SuperCOSMOS magnitudes were corrected using the Schlegel extinction maps
\citep{Schlegel98}. As a proxy for the specific star formation rate we take
the $\eta$ parameter as defined by \cite{Madgwick02} performing a principal
component analysis of the optical spectra. This parameter also shows a strong
correlation with the specific star formation rate obtained from the SDSS,
\citet{SolAlonso06}.  The parameter $\eta$ is correlated with the Hubble types
\citep{Norberg02}: $\eta < -1.4$ means mainly to E/S0-galaxies, while $ \eta >
-1.4$ corresponds mostly to late type S galaxies.  But the scatter is large
and we shall interpret our results in terms of the morphological type with
some care.

All galaxies with a quality $Q \ge 3$ were used in order to have accurate
redshift determinations. For determining the $e$- and $k$-corrections we
follow the method described in \cite{Folkes99} for those galaxies whose type
could be determined, and that by \cite{Norberg02} for those which could not be
determined and therefore was derived for a mix of galaxy types.  For the
SuperCOSMOS magnitudes we use the $k$-correction proposed by \citet{Cole05}
for the $r$-band magnitudes.

Eight volume limited samples where constructed both from coherent regions of
the SGP and NGP slices of the 2dFGRS. The SGP sample was restricted to the
right ascension range $ -2.^{\rm h}10 \le \alpha \le 3.^{\rm h}40 $, and a
declination range $ -33.^{\circ}00 \le \delta \le -25.^{\circ}50 $; the NGP
was selected through $ 10.^{\rm h}00 \le \alpha \le 14.^{\rm h}80 $, $
-4.^{\circ}00 \le \delta \le 1.^{\circ}00 $.

The sample characteristics are listed in Table \ref{vollim}. The samples N/S
1-4 were optimized for the void search, i.e. we set redshift cuts that provide
a large depth and a sufficient number of galaxies in the volume. In
particular, the selection window is chosen to be larger than typical void
sizes in order to minimize boundary effects in the void detection algorithm.
The samples have overlapping redshift ranges, but due to different faint
absolute magnitude limits $B_{J,{\rm lim}}$\footnote{Note that the $-5\log{h}$
  term is always included in our notation for absolute magnitudes $B_J$.}
voids in the different samples are defined by different galaxies.

The samples N/S 5-6 were constructed for investigating the properties of
galaxies inside of voids. Therefore, we have chosen exclusive magnitude
intervals $-18.4 < B_J < -19.2$ and $-19.2 < B_J < -20.075$ in the last four
columns of Table \ref{vollim}.

\begin{table}
\centering
\begin{tabular}{| l | r r r r l l |}
\hline
sample & $N_{\rm gal}$ & $B_{J,{\rm lim}}$ & $\lambda$ & $V/10^6$ &  
$z_{\rm min}$  &  $z_{\rm max}$   \\
 & & & \hMpc & $h^{-3}{\rm Mpc}^3$ &  & \\
\hline
N1 & 10609 & -18.0 &  3.82 &  0.59 & 0.014 & 0.086 \\
N2 & 16697 & -19.0 &  5.10 &  2.21 & 0.021 & 0.135 \\
N3 & 11749 & -20.0 &  8.27 &  6.64 & 0.033 & 0.198 \\
N4 &  1323 & -21.0 & 17.14 &  6.67 & 0.05  & 0.2   \\
S1 & 10343 & -18.0 &  4.69 &  1.07 & 0.014 & 0.092 \\
S2 & 18117 & -19.0 &  5.81 &  3.56 & 0.021 & 0.139 \\
S3 & 13964 & -20.0 &  9.00 & 10.20 & 0.033 & 0.2   \\
S4 &  1652 & -21.0 & 18.24 & 10.02 & 0.05  & 0.2   \\
\hline
N5 & 10389 & -18.4 &  4.25 &  0.80 & 0.05  & 0.1   \\
N6 &  8279 & -19.2 &  5.70 &  1.53 & 0.1   & 0.14  \\
S5 &  8175 & -18.4 &  5.28 &  1.20 & 0.05  & 0.1   \\
S6 &  9516 & -19.2 &  6.23 &  2.30 & 0.1   & 0.14  \\
\hline
\end{tabular}
\caption{Selection criteria for different volume limited samples used for void
  size statistics. Shown are the number of galaxies, the faint absolute magnitude 
  limit $B_{{J},{\rm lim}} $, the mean galaxy distance $\lambda$, the total
  volume $V$, and the minimum and maximum redshifts $z_{\rm min}$, $z_{\rm
  max}$ of the volume limited samples.}
\label{vollim}
\end{table}

\begin{figure}
  \epsfig{file=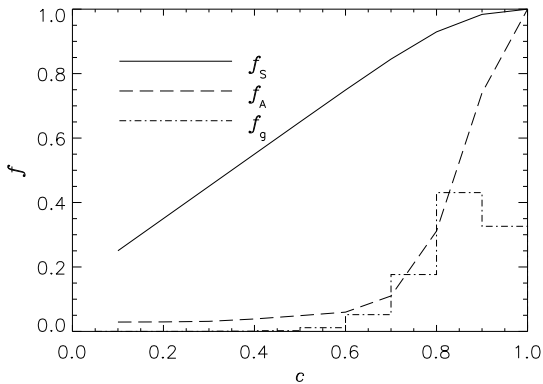,width=8cm}
 \caption{Fraction of the sky areas $f_A(<c)$ below a given completeness $c$ 
   (dashed line), differential fraction of galaxies $f_g(c)$ sampled with
   completeness $c$ (dash-dotted line) and galaxy fraction in the selected
   samples $f_S(c_l)$ diluted to a sampling fraction $c_l$ (solid line).}
\label{compl}
\end{figure}

In order to avoid biases in the void search due to an incomplete sky coverage,
in particular due to holes and missing fibers in the survey area, we take
special care in getting a homogeneously distributed completeness over the sky.
In Fig. \ref{compl} we show the fraction of the survey area $f_A(<c_l)$
covered by a completeness $c$ below a limit $c_l$ by the dashed line. The
fraction of galaxies contained in binned intervals of completeness $c$ are
given by the dash-dotted histogram $f_g(c)$.  Both distributions follow
directly from the survey completeness masks provided by \cite{Colless03}. To
get a homogeneously sampled set, we chose a completeness cut $c_l$, and we
reduce all fields with $c>c_l$ randomly to the fraction $c_l$. Then we get a
fraction in the galaxy samples $f_S(c_l)$ as

\begin{equation}
 f_S(c_l) = \int_{0}^{c_l} f_g(c) dc + 
                              \int_{c_l}^{1} f_g(c)[1-(c-c_l)]dc.  
\end{equation}

It is shown by the monotonously increasing solid line in Fig. \ref{compl}. We
do not care for that part of the sky covered only with completeness $c<c_l$.
If we chose $c_l = 0.6$ this concerns only about $5\%$ of the area.  On the
other hand, we have about $75 \%$ of galaxies in the samples $f_s(c_l)$. This
is a straight forward procedure that is adopted to our specific problem.

\section{Simulated galaxies}

For a quantitative comparison with CDM models of galaxy formation, we
compare the voids statistics and galaxy void properties with those
predicted by the semianalytical galaxy formation model of \cite{Croton06}.

This model uses the dark matter halo distribution and mass accretion histories
found in the Millenium simulation \citep{Springel05}, and applies a
semi-analytic formalism to describe the evolution of galaxies in these halos. 
The Millenium simulation covers a box of sidelength 500 \hMpc with 
$2160^3$ particles, run in a 
concordance cosmology ($\Omega_m = 0.25$ and $\sigma_8 = 0.9$). The large
volume is especially suited for the void search and for excluding boundary 
effects.

The key physical processes included in the semi-analytical model are 
shock heating of gas falling into the dark-matter halos, gas 
cooling and forming of galactic disks, star formation in the cold phase, 
supernova 
feedback heating of the cold phase, metal enrichment, and galaxy mergers
leading to spheroid formation, cp. \cite{cole00}. As new feature, `radio 
feedback' from AGN is implemented when a massive black hole forms at the 
centre of hot gas halo and suppresses further gas cooling. The radio feedback 
mode is needed in order to suppress star formation in massive dark 
matter halos. No feedback from quasar winds in included.

The model has been shown to reproduce the observed 2dFGRS galaxy luminosity
function, the Tully-Fisher relation, cold gas fraction/stellar mass and cold 
gas metallicity/stellar mass relations for Sb/c spirals, and the global color 
magnitude relation. It also reproduces the global star formation 
history of the Universe \citep{Croton06}.

The output catalog from the SAM model provides $B$, $V$  and $R$ band colors. 
We converted the $B$ band colors to $B_{J}$ using the relation given in
\citet{Norberg02}. We use four mock samples M1 to M4 with the same 
luminosity 
limits $B_{J,lim} = -18, -19, -20, -21$ as the volume limited samples for 
the 2dFGRS. In addition we select four mock samples MW1 to MW4 with 
the same magnitude limits and in addition with the window functions 
of the combined samples N/S1 to N/S4. In addition these samples are 
homogeneously 
diluted to a completeness of $c_{\rm l} = 0.6$. We do not impose the 
small incompleteness corrections for that part of the survey mask that
is less well sampled as 0.6. If not otherwise indicated the mock samples
are used in redshift space to allow a realistic comparison of the 
observed and theoretical void statistics.

\section{Finding Voids}

Although it is generally agreed upon that voids 
constitute underdense regions in the galaxy distribution, 
the precise definitions of voids 
varies considerably in the literature. Sometimes a 
threshold in the cosmic density field is imposed to define voids, 
either from the dark matter density, or constructing a continuous 
density field from the galaxy distribution. In either case, we 
must choose a smoothing scale to avoid strongly varying densities 
and corresponding irregular voids. A variant is the definition of 
wall galaxies from the local density field, cp. \citet{Elad97} and 
\citet{HoyleVogeley02}. 
Others define voids by regions empty of galaxies over a certain
magnitude \citep{Mueller00, Patiri06a} or of halos over a 
mass limit \citep{Gottloeber03}. 

In many previous void searches, 
the geometry of the voids is specified to be spherical, or a union of
overlapping spheres. Although from gravitational instability theory it 
follows that voids become more spherical during evolution as they expand
\citep{Icke84, Weygaert93}, 
in real data and during intermediate evolution stages,
they can have a wide range of geometries \citep{Shandarin04}.  
 
Voids can be seen as underdense perturbations in the initial Gaussian dark
matter density field that expand and become non-linear. Using the top-hat 
spherical collapse  (expansion) model, \citet{Sheth04} have shown that at 
the shell crossing time, the linear mean density should be $\delta = -2.41$ 
\citep{Sheth04}. In a recent paper \citet{Furlanetto06} have 
shown how this can be related to the mean galaxy density, using an Halo 
Occupation Distribution (HOD) model. Although this definition is appealing,
the corresponding galaxy density depends on the size of the smoothing length. 
It is therefore not obvious so far how to relate uniquely the galaxy
distribution to the underlying dark matter distribution. 

Therefore in this paper we define the voids as
regions being empty of objects larger than some given absolute magnitude 
$B_J$. As the more massive and brighter  galaxies will be found in 
filaments and clusters, we use them as tracers of the cosmic web. 
To investigate the self-similarity of the void distribution, we analyze 
a sequence of increasingly bright galaxies $B_J = -18, -19, -20, -21$, 
the samples N/S1-4 in Table \ref{vollim}. 

We used the void finder described in 
\cite[][]{Mueller00}. It is a grid-based void finder that locates the largest 
empty cube on the density grid. Next it looks for neighboring empty density
layers along all six sides of the cube. The largest compact empty layer 
consisting of a base square layer and extensions along the four square 
boundaries
is added to the void if its volume is a factor $f = 0.67$ larger than the 
previous boundary layer of the cube. 
This process is iterated along all six boundaries of the base 
cube, each time requiring that the additional layer is a boundary of the 
previous 
void boundary and exceeds this by the given factor $f$. This factor is also 
imposed in extending the base square layer along their for sides. This 
procedure provides almost convex empty voids. The value $f$ is the only free 
parameter in our void search algorithm. Its value is chosen so as to avoid 
narrow tunnels that go out of the base voids into the galaxy distribution. 
For the present analysis, we provide only results for the base square voids. 
We have analysed both 2dF-data and simulations also including the void 
search with extensions, with gives similar void size distributions, scaling 
relations and an reproduction of the data by the simulations. The reason for
restricting on the base voids lies in its use for looking for faint galaxies
laying in central regions of large voids.

For the analysis, a grid of resolution of 
1 \hMpc was chosen. The void finder is useful in that it is faster than looking 
for the largest empty spheres in the galaxy distribution. 
Although voids do not tend to be cubical, the extension along the boundaries 
and a sufficiently small grid allow for realistic void geometries. Comparison 
with the spherical void finder of \cite{Patiri06a} shows, that for large 
voids we are mainly interested in, the  void finder selects 
the identical void centers and sizes.

In our analysis we derive the size distribution of voids up to small values, 
thereby covering in most cases over 90~\% of the space with voids. For 
the comparison with other void finder, we use and effective radius 
$R = (3 V /4 \pi)^{1/3}$. 
As statistics we use the volume weighted void size distribution 
\begin{equation}
F(>R) = \int_R^\infty f(R) dR,
\end{equation}
defined as the cumulative 
volume of the survey covered by voids of radius larger than $R$. This definition
has the advantage of allowing a robust void size distribution over 2 to 3 order 
of magnitude in $F(>R)$. The strongly varying volumes of large voids then
appear at the low abundance end of the distribution. The more
common void probability function $P_0(R)$ is given by a weighted sum over the 
differential size distribution $f(R)= -dF/dR$, cp. \cite{Otto86, Betancort90}. 
The employed 
void size distribution $F(>R)$ does not suppose a pre-specified geometry 
of the voids, and it determines the {\it maximal} empty region in the galaxy 
distribution, starting from large voids and going to smaller ones. Thus it 
seems to be most sensitive to the large-scale distribution of galaxies 
in the cosmic web.   

\section{Results}

\subsection{The void size distribution}

First we investigate the void sizes found in the observed volume
limited samples. We found maximum voids with effective radii of 24 \hMpc 
and 22 \hMpc base sizes in the sample N4 and S4, resp. There are 
156, 70, and 2 voids over 
12 \hMpc base length in samples N4, 3, 2, and 1, resp.; and 
199, 116, and 2 voids in S4, 3, and 2, resp. (no such large void are found 
in N1 or S1). Voids over 6 \hMpc effective radius are much more abundant, there 
are 916, 1250, 321, and 66 voids in samples N4, 3, 2, and 1, resp.; and 
1364, 1786, 538, and 129 voids in samples S4, 3, 2, and 1, resp. 
Fig. \ref{voids2df} shows the 100 largest voids as found in the sample S3. 
For comparison Fig. \ref{voidsmill} shows the same for a Millenium mock catalog 
with the same magnitude range and an identical survey mask (i.e. MW3, see below). 
Obviously, both 2dFGRS- and mock samples look very similar, and they show
a comparable distribution of large voids. Large voids are more abundant at larger
distances from the observer. This is an effect of the survey window that
strongly restricts large nearby voids. Besides this geometric effect, large
voids are evenly distributed in data and model distributions, and void 
sizes are comparable in both samples. 

\begin{figure}
\epsfig{file=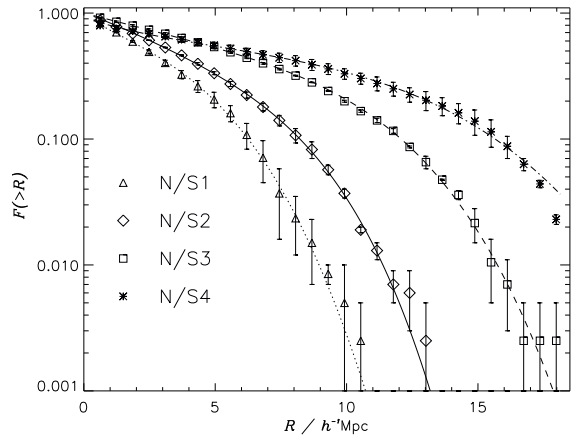,width=8cm}
\caption{Cumulative volume weighted void size distribution $F(>R)$ in the 
2dFGRS samples N/S1 (triangles), 2 (diamonds), 3 (squares) and 4 (stars). 
The analytic fit according to eq. (3) provide excellent descriptions of 
the data.}
\label{cum2df}
\end{figure}

More quantitative results can be seen in the cumulative void size distribution 
shown in Fig. \ref{cum2df}. There we combine the S/N samples from
the 2dFGRS to get better statistics. If taken separately we get similar 
curves for the north and south galactic pole regions. We take the difference 
for the separate distributions as a measure of the uncertainty in the 
samples shown by the error bars on the binned sample. 
Most obvious is the size dependence of voids on
the galaxy samples, the largest voids among bright galaxies with 
$B_J > 20, 21$ have an effective radius over 15 Mpc$/h$. 
The statistics of such large voids is still quite restricted, 
but smaller voids lead to size distribution with small error bars. 
The general form of the distribution is self-similar. We fit it with modified 
exponential distributions 
\begin{equation}
F(>R) = \exp\left[-\left({R\over s_1\lambda}\right)^{p_1} - 
            \left({R\over s_2\lambda}\right)^{p_2}\right]
\end{equation}
with 4 parameters, two length factors $s_1$ and $s_2$, and two powers 
$p_1$ and $p_2$. The fits are by $\chi^2$-minimization and always provide
excellent representations of the observational data covering more than 
2 orders of magnitude in $F(>R)$. The main scaling of the galaxy samples 
is described by the mean separation between galaxies, the scale $\lambda$. 
This value and all fit parameters are given in Table \ref{fit}. 
The first exponential describes the statistics of 
small and intermediate sized voids. This part is typically described
with an linear fall-off in the exponential function, i.e. $p_1 \approx 1$. 
For describing the cutoff of the void size distribution for large void radii, 
we need the second factor in the exponential distribution with a higher
power of the void radius, typically with $p_2 \approx 3$, i.e. the abundance
of large voids is cut off with the void volume entering the exponential
function. The length factors $s_1$ and $s_2$ are both 
of order unity. After multiplication with the mean galaxy separation 
they provide the void radii where the two exponential laws dominate the 
size distribution. There are no ways to fit the size distribution with 
a single exponential distribution.

\begin{figure}
\epsfig{file=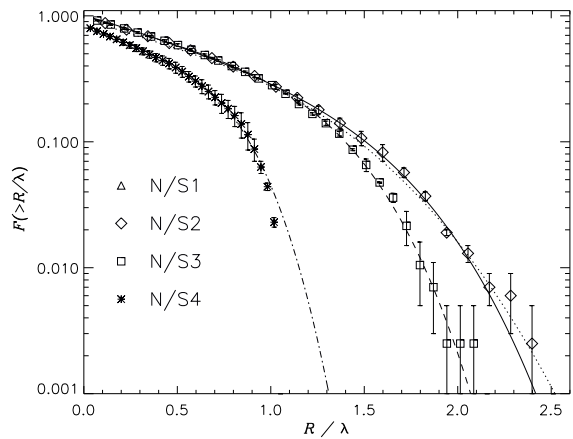,width=8cm}
\caption{Normalized void size distribution $F(>D/\lambda)$ as function of the
void diameter divided by the galaxy separation $\lambda$ in the 2dFGRS with 
same line symbols for samples N/S1-4 as in Fig. \ref{cum2df}.} 
\label{cum2dfnorm}
\end{figure}

In Fig. \ref{cum2dfnorm} 
we show the normalized void size distribution of the four 2dFGRS data
sets, i.e. we divide the void sizes $R$ by the mean separation between 
galaxies $\lambda$.  Obviously, the void size distributions are self-similar in this
variable. The scaled void size distributions coincide at radii 
$R < 1.5\lambda$. This scaling is similar to the 
dependence of the correlation length of galaxies and groups 
as found by \citet{Bahcall92} and \citet{Yang05}. 
Note that the samples N/S4 are 
strongly restricted by the survey geometry. The mean intergalaxy separation 
of about 18 \hMpc is comparable to the thickness of the 2dFGRS slices 
of about 45 \hMpc at the far side of the selected volume, 
even if the length of the slices much exceeds the 
void sizes. Effectively the statistics of these voids, shown by the 
dash-dotted lines in Fig. \ref{cum2dfnorm}, is cut off at an volume 
filling fraction of 2 \%. Most interesting in the scaled
void size distribution is that the exponential cutoff in the 
void size distribution depends on the galaxy sample defining the 
voids. Voids among fainter galaxies have a larger tail in the 
distribution of void radius over mean galaxy separation, $R/\lambda$. 
In this variable, the largest voids are among the 
{\sl faint galaxy samples} N/P1 and 2. 
Generally the void radii $R/\lambda$ extend over larger ranges than 
voids in Poisson samples. The voids are larger than 
in a random galaxy distribution, and the size distribution shows a sharp 
cut-off what is an indication the the voids are hitting the galaxy walls 
in the large cosmic web. 
 
\begin{table}
\centering
\begin{tabular}{| l |  r r r r l l |}
\hline
catalog & $\lambda$ & $p_1$ & $p_2$ &  $s_1$ & $s_2$ \\
        & ${\rm Mpc}/h$ &   &  & & & \\
\hline
N/S1 &  4.26 & 0.8 & 3.0 & 1.02 & 1.49 \\
N/S2 &  5.45 & 1.1 & 4.5 & 0.86 & 1.80 \\
N/S3 &  8.64 & 1.0 & 4.5 & 0.93 & 1.47 \\
N/S4 & 17.69 & 0.6 & 4.0 & 0.67 & 0.86 \\
\hline
MW1  &  4.16 & 0.7 & 2.0 & 1.70 & 1.26 \\
MW2  &  5.55 & 0.8 & 3.0 & 1.10 & 1.50 \\
MW3  &  9.13 & 1.0 & 4.0 & 0.86 & 1.46 \\
MW4  & 22.21 & 0.5 & 4.0 & 0.43 & 0.70 \\
\hline
M1   &  3.58 & 0.9 & 3.0 & 1.36 & 2.07 \\
M2   &  4.83 & 1.0 & 3.5 & 1.23 & 2.02 \\
M3   &  8.02 & 1.1 & 4.0 & 1.18 & 1.71 \\
M4   & 19.60 & 0.6 & 3.5 & 3.79 & 1.17 \\
\hline

\end{tabular}
\caption{Fit parameters of void statistics for volume limited samples 
  of the 2dFGRS and of galaxy samples (NGP and SGP) and in the Milleninum 
  simulation (MW with window selection from 2dFGRS, M for the complete box). 
  The different samples are characterized by the mean galaxy distance 
  $\lambda$.}
\label{fit}
\end{table}

\begin{figure}
\epsfig{file=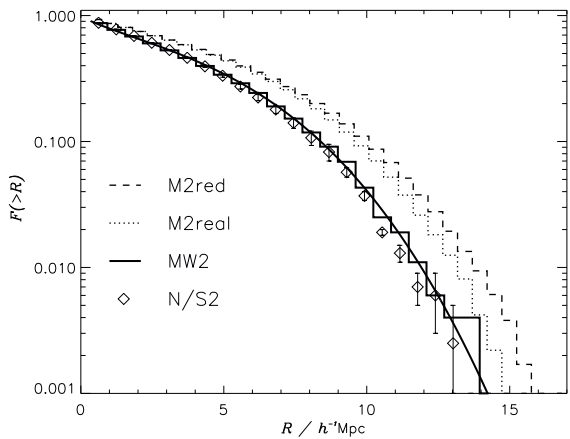,width=8cm}
\caption{Comparison of the S/N2 void size distribution of the 2dFGRS 
with the data from the Millenium mock catalogue MW2 modelling the 
selection window of the observations (thick histogram and solid fit curve). 
The dashed and dotted lines are the void size distributions of the complete
Millenium box in redshift and real space, resp.} 
\label{cum2dfmill}
\end{figure}

We compare the void size distribution of the 2dFGRS with the mock samples 
from the Millenium simulation in Fig. \ref{cum2dfmill}. Here we show the 
samples N/S2 with the diamond symbols and error bars. The solid histograms 
and the fit from Table \ref{fit} shows the results for the Millenium galaxy 
mock sample restricted to the 2dFGRS survey window (the sample MW2) 
with $B_J<-19$. In addition we show the void statistics for the complete
Millenium volume (sample M2) in real (dotted line) and redshift space 
(dashed line). These two distributions are quite similar, but systematically 
shifted to larger void radii by about 5\% larger in redshift than in real 
space. This is an effect of the overall expansion of the underdense regions 
in comoving coordinates, i.e. forground and background void boundaries 
show systematic deviations from the Hubble flow. The 
mock sample MW2 in the 2dFGRS-window is shown in redshift space. Within 
the error bares, it well reproduces the void statistics of the N/S2 sample. 
There is a slight tendency of overpredicting the abundance of large voids 
in the simulations. It should be noted that the void sizes $R$ in the 2dFGRS 
survey window are nearly 20\% smaller than in the 500 \hMpc. 
Still for the 2dFGRS window geometry, finite volume effects
are important for all the range of the void statistics. The general agreement
of the model and 2dFGRS void statistics holds true for the other 
data N/S1, 3, and 4; and the mock samples MW1, 3, 4, as the parameters 
of the fit curves in Table \ref{fit} demonstrate.

\begin{figure}
\epsfig{file=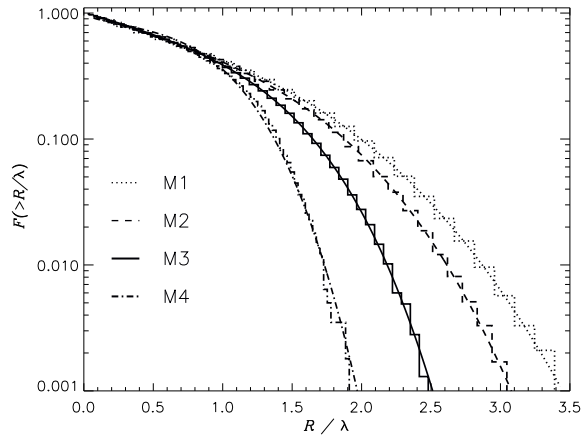,width=8cm}
\caption{Normalized void size distribution in the complete Millenium box for 
 magnitude limits $B_J < -18, -19, -20, -21$, i.e. the samples M1-4 from
 right to the left.}
\label{cummill_norm}
\end{figure}

Most interesting is the dependence of the void size distribution on the
limiting magnitude, shown in Fig. \ref{cummill_norm} for the full statistics
of the Millenium box. This is the complement to Fig. \ref{cum2dfnorm}. 
Below void radii $R < 1.5 \lambda$, the void size
distributions coincide for different samples. 
This means void sizes depend strongly on the 
mean galaxy separation, and for a long range 
they are proportional to $\lambda$. Beyond this radius, 
$R > 1.5 \lambda$, void sizes depend strongly on the magnitude cut. Voids
among faint galaxies have an more than three times larger radius $R$ than the mean 
intergalaxy separation, see the results for $B_J > -18$ galaxies 
(dotted histogram). In the contrary, voids among
bright galaxies with $B_J > -21$ have radii only up to $2 \lambda$. This
is an expression of \citet{Peebles01} void phenomenon: There are relatively 
large voids among faint galaxies. And otherwise, voids among bright galaxies 
are also underdense in faint galaxies which are found predominantly at 
in the outer regions of the voids from bright galaxies.  

\begin{figure}
\epsfig{file=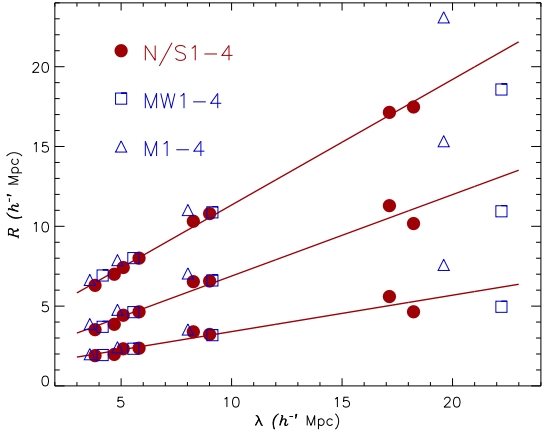,width=8cm}
\caption{Median and quartiles of the volume covered by voids as function of 
the mean intergalaxy separation. The full dots 
stem from 2dfGRS data, the squares from mock samples MW1-4, and the triangles 
from the complete Millenium box samples M1-4. The lower
25\% percentile, the medium and the upper 75\% percentile are described by 
linear fits, curves from below.} 
\label{median}
\end{figure}

The self-similarity of the void size distribution that is seen in the 
proportionality of the void size distribution to the mean intergalaxy
separation becomes also obvious in the radii of voids that cover 25 \%, 
50 \%, and 75 \% of the space, i.e. the median and the quartiles of 
the void size distribution of Fig. \ref{cum2df}. The filled squares in 
Fig. \ref{median} show these percentiles of the N1-4 and S1-4 volume
limited samples of the 2dFGRS. The differences in the mean galaxy
separations and void percentiles between neighboring data points show the 
uncertainty due to cosmic variance. They are of the same order as 
the symbol sizes. These percentiles are well described by linear fits 
to the lower 25\% quartile $R_{25}$, the median $R_{50}$, , and the upper 
75\% quartile $R_{75}$, 
\begin{equation}
 R_{per} = R_{0} + \nu \times \lambda, 
\end{equation}
where an initial value $R_0 = 1.1 \pm .2, 1.8 \pm .4, 3.5 \pm .2$ \hMpc 
and a slope $\nu =0.23 \pm .02, 0.51 \pm .04, 0.79 \pm .01$ for the 25\%-, 
medium and 75\%-percentiles are found by linear regression of the data. 
The squares correspond to the mock samples 
MW1-4 from the Millenium galaxy catalogues with limiting magnitudes of 
$B < -18, -19, -20, \rm{and} -21$. For the 3 fainter samples, they 
fall completely on the observed 
similarity relations. The triangles are the samples from the full 
Millenium box M1-4. For samples with $\lambda < 10 \hMpc$ they lie 
about 10 percent above the data and simulation data in the 2dFGRS 
window. These samples are not influenced by boundary effects. 
The large voids for $\lambda > 15 \hMpc$ show even larger differences
between the samples in the simulation box and in the window volume.
This similarity relations were previously found in the 
LCRS void analysis 
\citep{Mueller00} and in CDM simulations \citep{Benson03} but there with
less statistics, i.e. larger uncertainties. 

\subsection{Properties of void galaxies}

Now we describe our results concerning the void galaxies. We ask whether
galaxies with magnitudes fainter than the lower absolute magnitude limit
$B_{J,\rm{lim}}$ of the volume limited samples for the void search are
populating the voids and whether their properties differ from average
galaxies.  Because by definition the voids are sparsely populated we stack all
voids over a minimum void size $R_V \ge R_{V lim}$ and look for all void
galaxies within a scaled radius $f_R=R/R_V$. We investigate voids identified
in the volume limited samples N/S5 and 6 with magnitude limits $ B_J \in
\left[-19.2,-18.4\right]$ and $ B_J \in \left[-20.075,-19.2\right]$. We take
all voids larger than the limiting diameter $R_{V lim} = 10 \hMpc$ for the
analysis.  To attain better number statistics we include all galaxies in the
detected voids, also those that had been removed while correcting for
nonuniform completeness during the analysis of the void size statistics.  In
this way we obtain a sample of 661 galaxies in the fainter and 1222 galaxies
in the bright sample.  Although the luminosity function is a decreasing
function with increasing luminosity, we obtain more galaxies in the brighter
sample, because their brightness allows a deeper redshift range and a larger
volume (cp.  Table \ref{vollim}). We choose to analyze separately the central
parts of voids with $f_R = 0.3$ and the overall properties of void galaxies
with $f_R = 0.6$.

\begin{figure}
  \epsfig{file=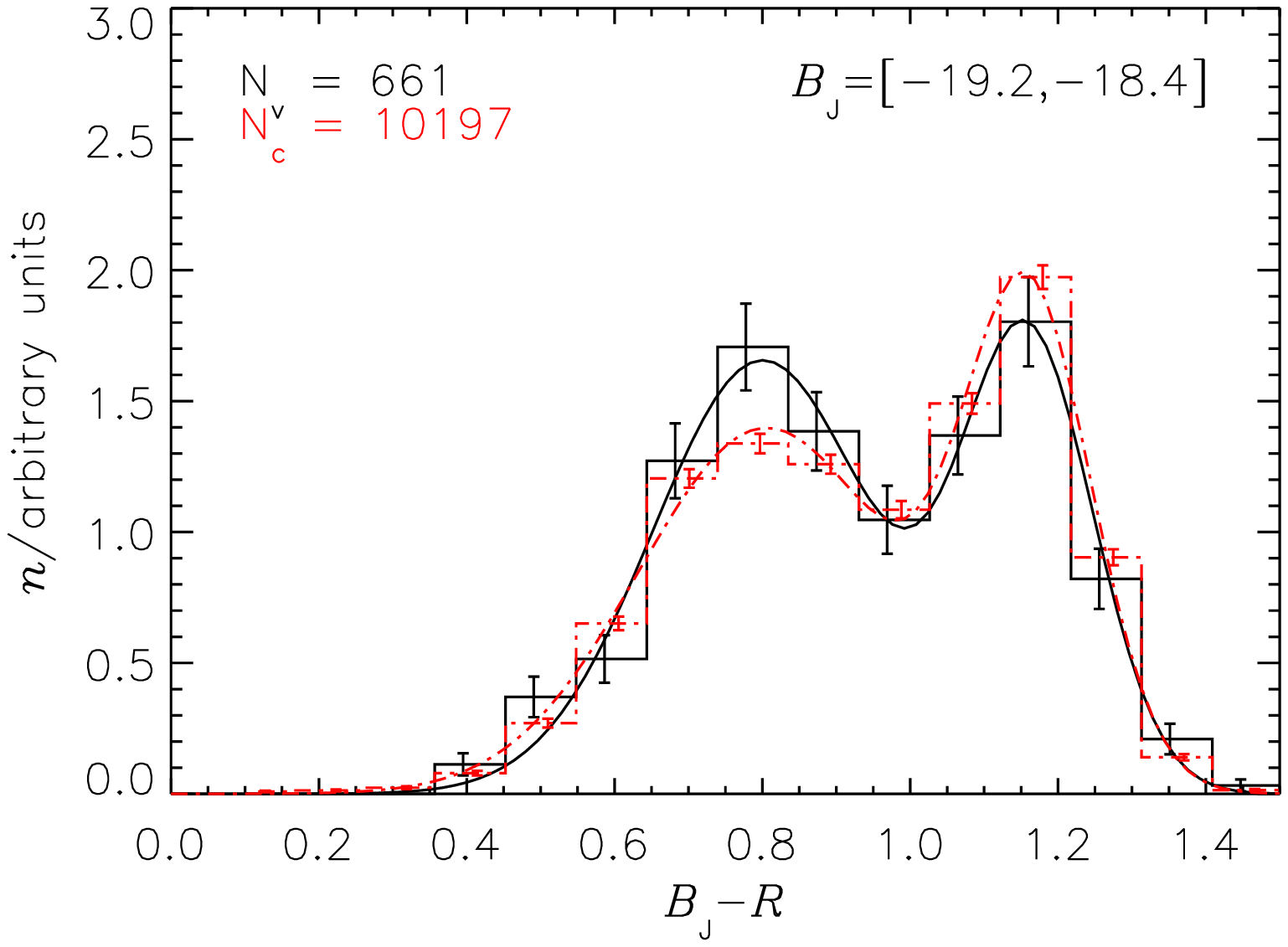,width=8cm}
  \epsfig{file=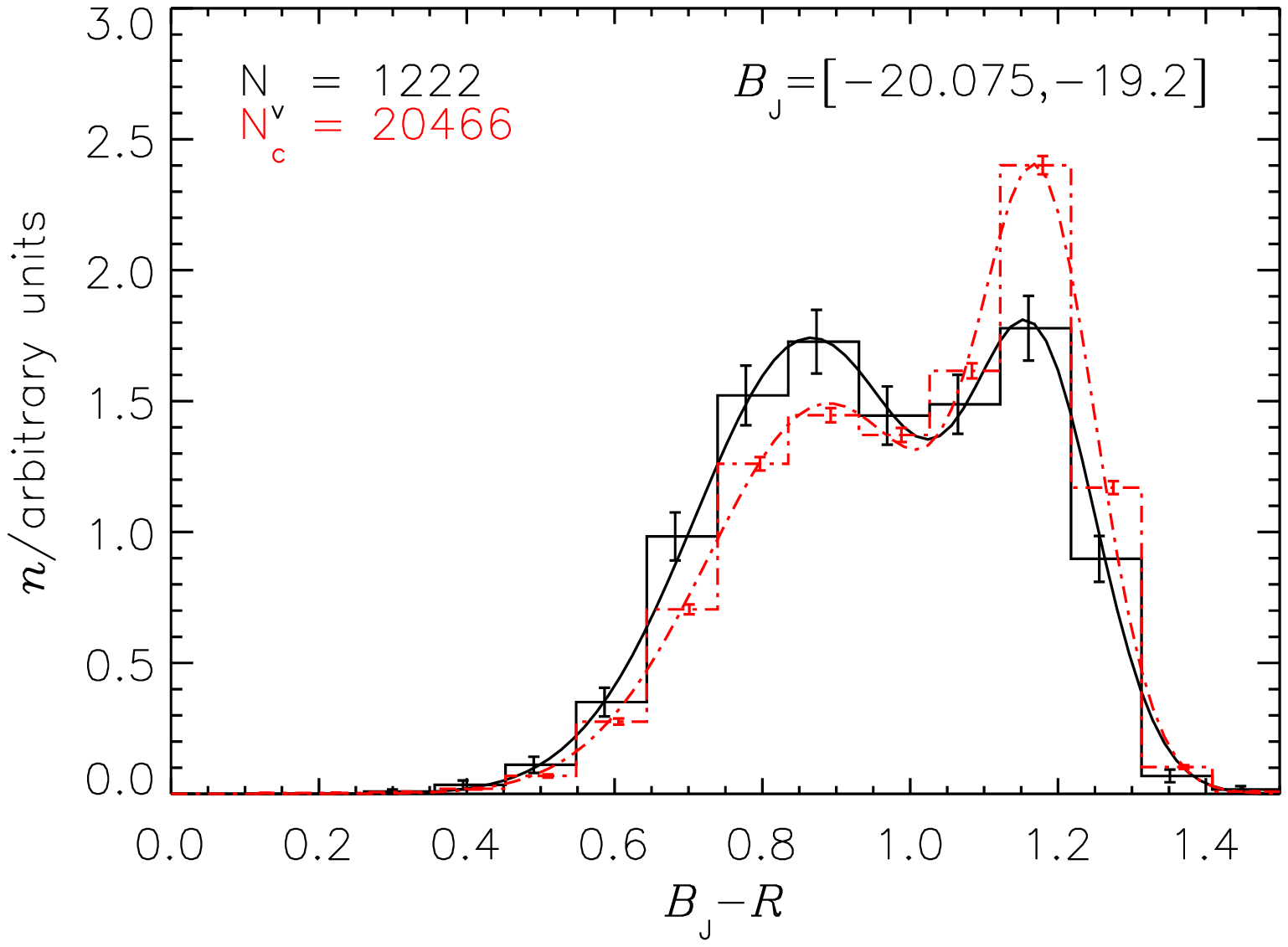,width=8cm}
 \caption{The magnitude dependence of the ($B_{J}-R$) color distribution 
   of galaxies in void centers $R/R_v \le 0.6$ (solid histogram with $1\sigma$
   Poisson error bars) and double Gaussian fit (solid line). {\it Top}: The
   fainter and {\it bottom} the brighter magnitude ranges given in the upper
   right corner of both panels). The dash-dotted lines are the control samples
   in the complete survey volume. The galaxy numbers are shown in the upper
   left corners.}
\label{magdepend}
\end{figure}

In Fig. \ref{magdepend} we show the normalized color distributions of void
galaxies within $f_{\rm R} = 0.6$ as histograms with Poisson errors together
with double Gaussian fits obtained with $\chi^2$-minimization as solid
histograms and lines, respectively.  As control sample for average galaxies,
we randomly choose an equal number of spherical regions in the 2dFGRS volume
as the number of voids.  The number of galaxies is given in the upper part of
the diagrams.  We found a clear bimodality both in the void galaxies and in
the control sample for both brightness bins (taken NGP and SGP samples
together). This is the first time that this color bimodality has been shown
for void galaxies in the 2dFGRS. Both for the brighter (lower panel) and
fainter (upper panel) samples, inside voids the fraction of blue galaxies
increases on account of the fraction of red galaxies. In as far as the color
is a measure of the morphology of the galaxies, the changed proportion of blue
and red galaxies is a continuation of the morphology density relationship of
cluster galaxies by \citet{Dressler80} and discussed by \citet{Hogg04} for
SDSS galaxies estimating the density on a local scale of 1 \hMpc projected on
the sky and by 1000 km/s in radial direction. Fig. \ref{magdepend} shows that
the suppression of the red sequence inside voids is stronger for the brighter
galaxies. The ratio of void galaxies in the red and blue clouds are almost
independently of the brightness, while the field population has a stronger red
sequence population in the brighter bin, i.e. there is a stronger effect of
star formation strangulation than in voids.
 
\begin{figure}
  \epsfig{file=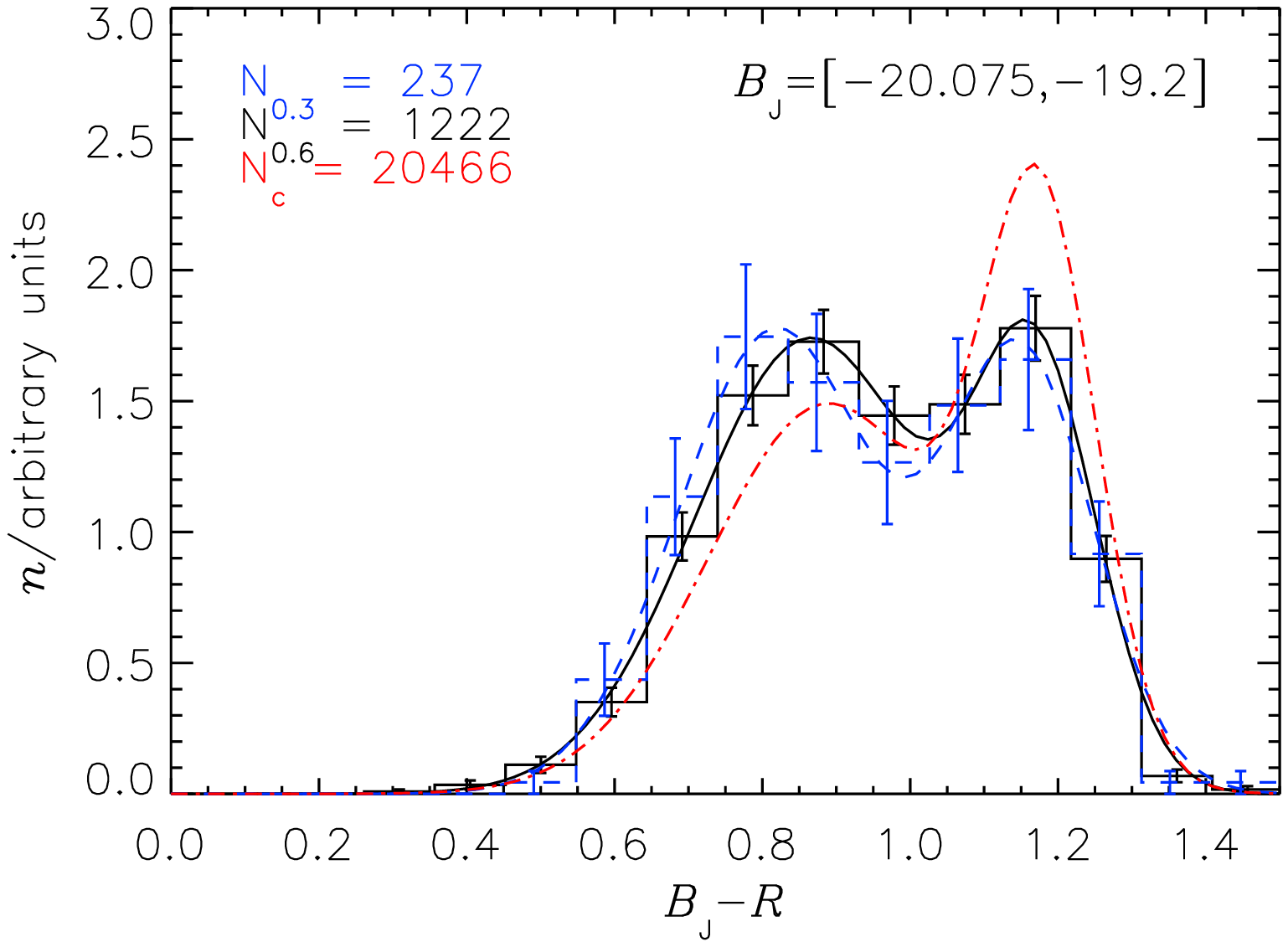,width=8cm}
 \caption{The radial dependence of the ($B_{J}-R$) color distribution of the 
   brighter void samples. The solid distributions show the $R/R_v \le 0.6$ and
   the dashed distributions the central $R/R_v \le 0.3$ void regions.  The
   dash-dotted line repeats the comparison sample, the galaxy numbers are
   again given in the upper left corner.}
\label{raddep}
\end{figure}

For the brighter galaxies, the color distribution of the void sample has been
analyzed for different fractions of the void size, $f_R = 0.3, 0.6$
(Fig.~\ref{raddep}). It is interesting that the blue cloud seems to show a
blue shift by about 0.1 mag, while the red sequence remains fixed at about
$B-R=1.2$.  So there is a weak tendency that the blue galaxies in most
underdense regions are not only more abundant, but also tend to be in average
{\it bluer} and maybe younger than galaxies in the same magnitude range in the
control sample.  The faint bin does not contain enough galaxies to plot the
color distribution for more centrally confined galaxies.

\begin{figure}
  \epsfig{file=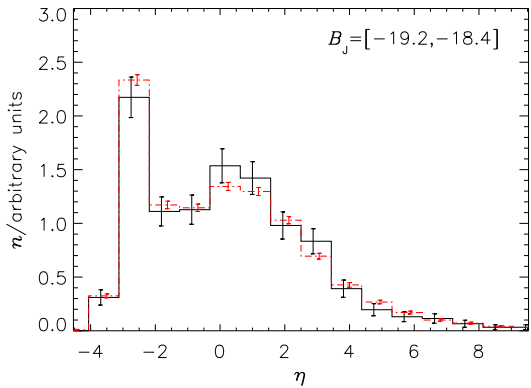 ,width=8cm}
  \epsfig{file=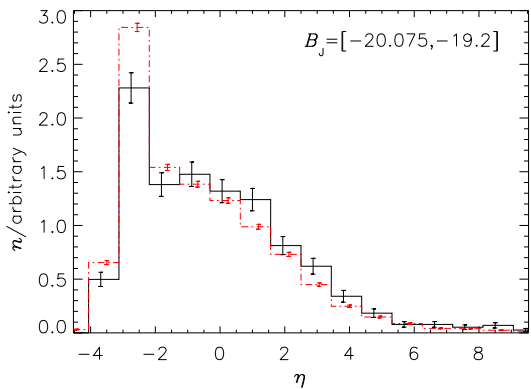 ,width=8cm}
 \caption{{\it Top}:  Distribution of the spectral parameter $\eta$ 
   in voids ($R/R_v \le 0.6$) (solid lines) and in void centers $R/R_v \le
   0.3$ (dashed lines), the control sample is given by the dash-dotted line.
   {\it Top:} The faint and {\it bottom} the bright galaxy sample.}
\label{sfrdist}
\end{figure}

For the same selection criteria we show the distribution of the $\eta$
parameter as a proxy for the specific star formation rate in
Fig.\ref{sfrdist}. For both the bright and less bright sample, the star
formation rate of void galaxies is enhanced, although this is only marginally
significant for the lower magnitude bin.

\begin{figure}
  \epsfig{file=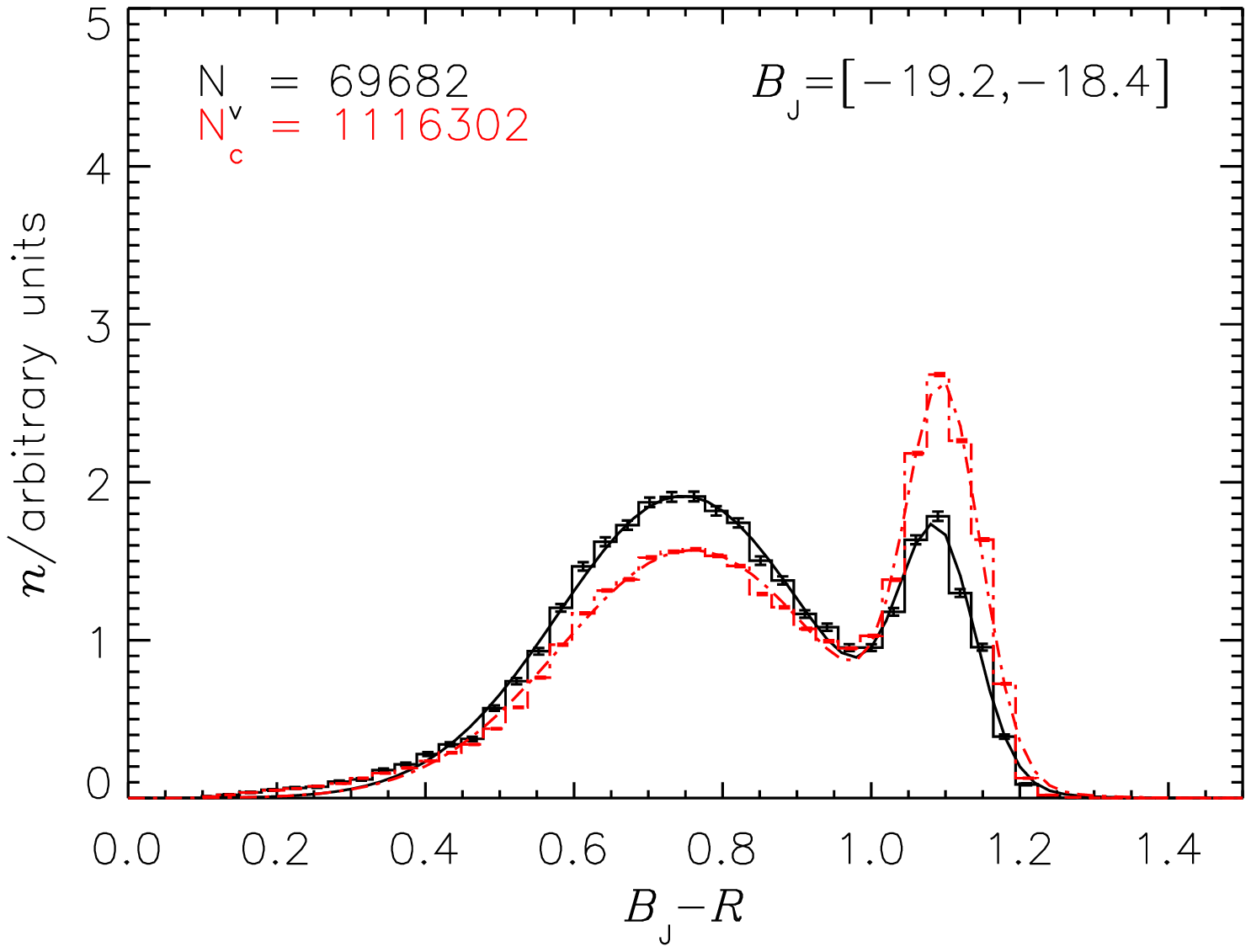,width=8cm}
  \epsfig{file=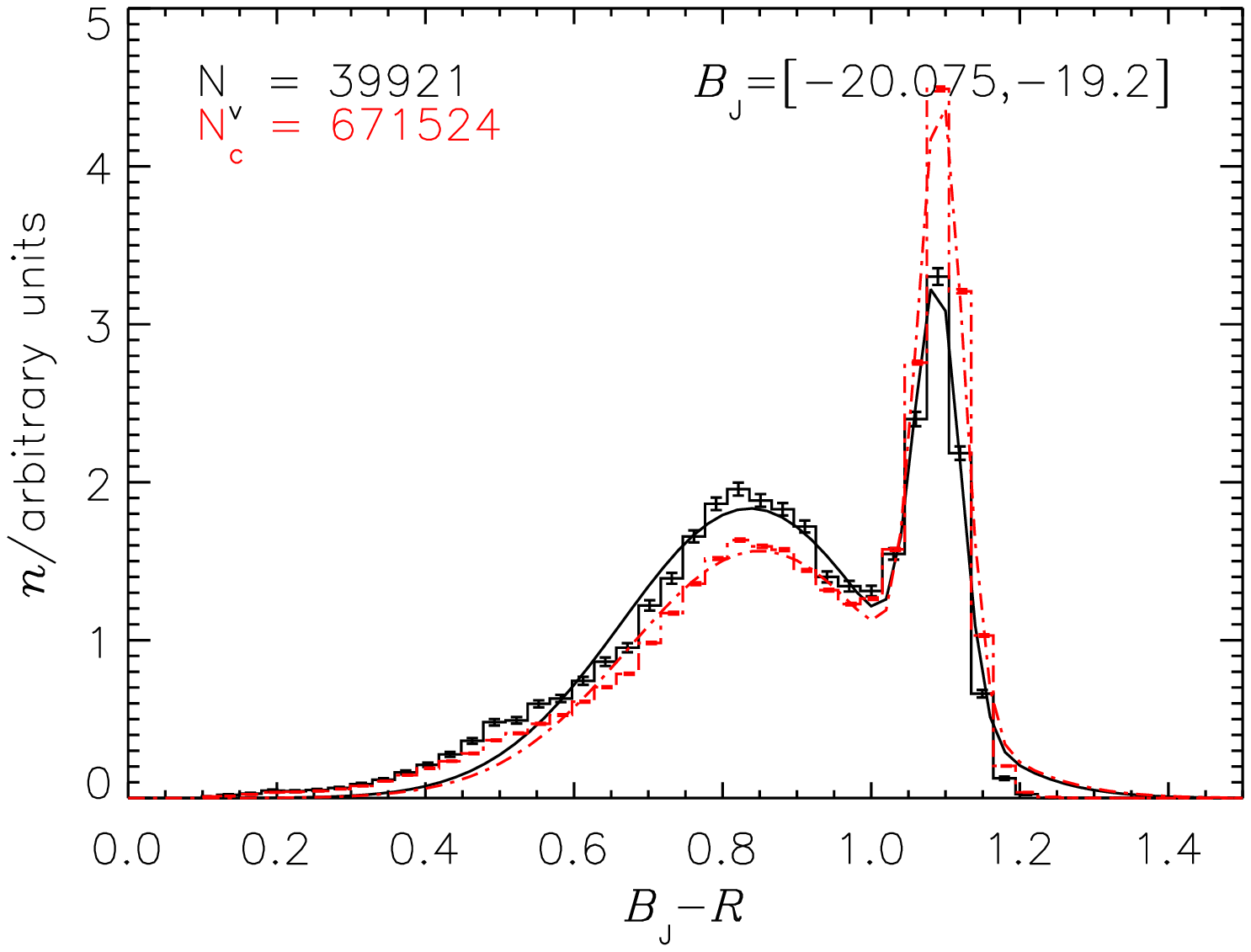,width=8cm}
 \caption{The magnitude dependence of the color distribution $(B_J-R)$  
   of void galaxies in the Millenium mock galaxy sample. The solid
   distributions show the void and the dash-dotted distributions the control
   samples. {\it Top}: The faint and {\it bottom} the faint galaxy sample.}
\label{magdepend_mil}
\end{figure}

From the semi-analytical model in the Millenium simulation we get
qualitatively similar color distributions of the void galaxies. We show in
Fig. \ref{magdepend_mil} the $(B_J-R)$ color distribution of the faint (above)
and bright galaxies (below). Again the distributions can be well fitted with
double Gaussian distributions. In the model samples, the mean colors of both
blue and red galaxies are the same for voids and the field, i.e. the void
environment has an influence on the relative abundance of the both galaxy
species, but not on its colors.  Comparing Fig. \ref{magdepend} and Fig.
\ref{magdepend_mil} we denote some quantitative differences. The suppression
of the red peak inside voids is more pronounced for the faint magnitude bin,
and altogether the red galaxies have a much tighter color range. The peak of
the red sequence of the brighter galaxies is much sharper than in the observed
sample, and it lies at a slightly less red color, $B_J-R \approx 1.1$. So
although the location and the width of the blue cloud is in good agreement
with the data, the semianalytical models have too many red galaxies with a too
sharp color distribution.  Furthermore, the magnitude dependence of the ratio
of red and blue galaxies shows significant differences to the data. The faint
galaxies show the stronger suppression of the red sequence than the brighter
magnitude bin. A similar discrepancy of the fraction of red galaxies inside
galaxy groups of the SDSS redshift survey against the Millenium simulation
semianalytic galaxy catalogues has been found by \citet{Weinmann06}.

\begin{figure}
  \epsfig{file=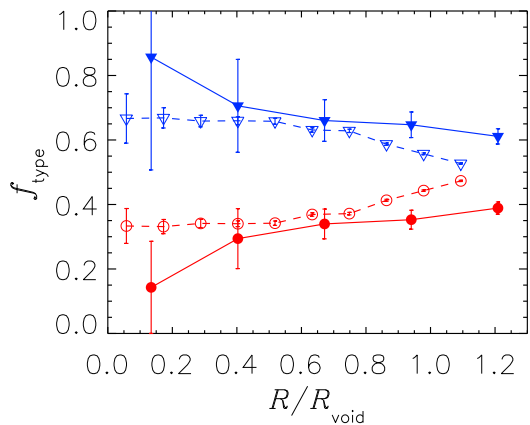,width=9cm}
 \caption{Radial distribution of the fraction $f_{\rm type}$ of early (E, lower spherical
   symbols) and and late (S, upper triangle symbols) in dependence of the
   scaled void radius. Solid symbols stem from the 2dFGRS and open symbols
   from the Millenium mock galaxy sample both in the redshift range $-18.4 >
   B_J > -20.075$.}
\label{rad_colfrac}
\end{figure}

Finally, we compare the radial distribution of the fraction of late (S) and
and early type (E) void galaxies in the 2dFGRS and in the Millenium simulation
in Fig. \ref{rad_colfrac}. In the 2dFGRS we employ the spectral parameter
$\eta$ for roughly distinguishing early and late galaxy types, with $\eta \le
-1.4$ for E and $\eta > -1.4$ for S galaxies. In the semianalytical model, S
galaxies are defined by a mean star formation rate $\log SFR/M_{\rm sun}/{\rm
  yr} > 0$, in the contrary E galaxies by $\log SFR/M_{\rm sun}/{\rm yr} \le
0$. These definitions lead to a fraction $f=0.52$ of E-type galaxies both in
the 2dFGRS and the Millenium catalogue. We show the fraction of E- and
S-galaxies within voids of radius larger then 13 \hMpc and for the combined
samples N/S 5 and 6, i.e. the range $-18.4 < B_J <-20.075$, solid symbols for
the 2dFGRS and open symbols for the Millenium galaxies.  Obviously both
distributions are quite similar. They coincide within the $1\sigma$ error bars
at mean radii values but the 2dFGRS shows a stronger radial variation of the
spiral and elliptical fraction There is the general tendency that the fraction
of E-galaxies in observed voids is smaller than in the simulations, this
tendency extends over the total radial range of voids. Furthermore, in the
2dFGRS, the difference in the E/S-fraction extends beyond the void volume,
i.e. the differences in the star formation activity is more pronounced and
extends over larger scales in the data than in the comparison semianalytical
model.

\section{Discussion}

\subsection{Void size distribution}

We found a clear demonstration of the similarity relation of void sizes in the 
2dFGRS, $R_{\rm{void}} \propto \lambda$, first established empirically both in 
observations and in models \citep{Mueller00}. There the much smaller LCRS was
analyzed and the analysis was done in 2-dimensional slices due to the geometry 
of this survey. Later similar relations were found for voids in the dark matter
density field using the void probability function \citep{Schmidt01}, among CDM-halo
catalogues \citep{Arbabi02}, and among semianalytical galaxy catalogues
\citep{Benson03}. Here we could provide an analytical description of the void size
distribution given by Eq. (3). The similarity relation Eq. (4) is a direct
consequence of such a void size distribution. The scaling of the median sizes
is comparable to our earlier findings taking into account the transition to the
3-dimensional survey geometry. 

The scaling relation (i.e. the constant of proportionality) of void sizes 
$R \propto \lambda$ is flatter than found by \citet{Benson03}. We suspect this
is due to our procedure of identifying both large and small voids with the
same algorithm, not just looking inside of wall galaxies of fixed local
overdensity as other void finders do. 
The scaling relation is a representation of the self-similarity 
of the void phenomenon. In 
distinction to earlier analyses, we have a large sample of voids and
therefore a good statistics in determining the void size distribution. 
The flat scaling relation is a consequence of relatively small voids among 
the bright galaxy samples with large $\lambda$ and conversely, 
comparatively large voids among faint galaxies. This is a new property well 
established  in the densely sampled 2dFGRS and in the large Millenium
simulation. It is the same void phenomenon discussed 
in the stimulating paper by \citet{Peebles01}. However, he 
claims a significant discrepancy of the void phenomenon and the standard 
hierarchical clustering phenomenon. Contrary to this claim we established 
void statistics that are well reproduced by the Millenium galaxy
catalogues. As shown in Fig. \ref{cummill_norm} the size distribution 
of voids in simulated galaxies is well described by a self-similar 
distribution for small and intermediate sized voids, and a more extended tail 
of large relative void sizes with respect to the mean galaxy separation. This means 
that the void phenomenon discussed by \citet{Peebles01} is a 
genuine consequence of the underlying LCDM model. An earlier 
demonstration of the void phenomenon within the hierarchical CDM clustering
model by quite different methods was provided by \citet{Mathis02}.

The new results concern the precise verification of the void scaling relations for
the 2dFGRS survey and the detailed fitting of the void size distribution by a
two-branch exponential distribution. Both properties of the 2dFGRS are
surprisingly well reproduced in the semianalytical galaxy formation scheme of
the Millenium simulation \citep{Croton05}. The comparison is done in the 
2dFGRS survey mask and in redshift space. This is essential for the precision 
comparison since we found coinciding void size distributions both for 
small and median void sizes, but an increase of the largest voids in redshift 
space by about 10\%, and an overall reduction of the void size distribution 
in the 2dFGRS survey mask as compared to the complete 500 \hMpc simulation box.

\subsection{Void galaxies}

We established a clear representation of the color bimodality of void galaxies 
and an larger fraction of blue galaxies in the 2dFGRS voids. 
An increase of the fraction of blue galaxies in voids has been previously 
established by \citet{Hoyle04} for the 2dFGRS and by \citet{Rojas04, Rojas05} 
for the SDSS. Also, \citet{Croton05} found a dependence of the blue fraction 
on the large-scale density environment of a scale of $8 \hMpc$. 
\citet{Kauffmann04} investigated the SDSS for the environment dependence of 
star formation activity of galaxies on a more local scale. They found that the 
most sensitive environmental dependence of galaxies is the SFR, which is
strongest 
for galaxies with stellar masses $< 3 \times 10^{10} M_{\sun}$, but no much 
environmental dependence on the large scale density was found when the small
scale density was specified. 

A bimodality of the color distribution of void galaxies in the SDSS 
has been established 
by \citet{Patiri06a}. Our results for the 2dFGRS are new and extend 
these earlier findings. As in that paper, we confirm the 
stability of the red sequence in void regions, but we find a slight blue 
shift of the blue cloud. Furthermore, there is a weak 
tendency of void galaxies to have a higher SFR, and for more early type
(E) than late type (S) galaxies. A reason for the blue slight shift of the 
blue cloud established here but not by \citet{Patiri06a} in the 
SDSS analysis may be that we analyzed the central void 
regions, while they considered the whole void volume. 
Also the 2dFGRS is denser sampled, and therefore we have a better statistics 
of void galaxies. While 
the SDSS colors have a better photometric accuracy, it was shown  
by \citet{Norberg02} that there is only a small non-linearity between 
the APM and SDSS colors. Some difference might be produced by differences in 
the adopted $k$- and $e$-corrections. 

Color distributions at higher redshifts show a similar bimodal behavior 
as our results. By comparing the SDSS color distribution with DEEP2 results, 
\cite{Blanton06} found that the blue
cloud is shifted bluewards by 0.6 mag at redshift $z=1$. 
Of course, this shift is stronger   
than what we find here, but it indicates 
a similar trend. Therefore a natural interpretation would be a slower 
evolution of structures inside voids and a younger 
galaxy population as compared to the field.

If the color shift might be due to a younger population in underdense regions,
we would expect the shift in the blue peak to be stronger for the lower magnitude 
bin. However due to the low number of void galaxies in the fainter magnitude
bin, we cannot establish such a tendency. 

The color shift is of similar amount than the color uncertainty of 
$\delta m=0.09$. But an uncertainty would basically increase the broadness
of the blue cloud galaxy distribution, and not leading to a systematic
shift. Therefore we conclude that the precise color distribution of blue 
void galaxies should be further investigated.

\section{Conclusions}

In this paper we have studied properties of voids in the 2dFGRS and
compared them with those predicted by semi-analytical models of galaxy 
formation. In particular, we were interested in the distribution of void 
sizes in the galaxy distribution as a function of the faint limiting magnitude 
of the sample. We established an almost linear dependence of the void 
size distribution on the mean separation $\lambda$ of the galaxies in the 
sample. A similar relation was found for the magnitude dependence 
of the correlation 
length of galaxies and galaxy groups \citep{Bahcall92, Yang05}. 
It is a natural consequence of the halo model of gravitational 
clustering and the resulting void statistics \citep{Tinker06}. 

In addition to the self-similarity, 
we have found that the voids among faint galaxies extend 
to relatively larger scales when devided by the mean void size. 
In the scaled radius $R/\lambda$, the 
voids size distribution of faint galaxies has a longer tail at 
large radii than those for brighter galaxies. This indicates 
that the faint galaxies trace the general spatial distribution of the 
cosmic web of brighter galaxies. 

For the 2dFGRS, we find more large voids 
at larger distances from the observer. This is due to the conelike 
structure of the survey. In our mock sample, we use exactly the 
same geometry of the observed samples. Thereby we 
found that the number density of the largest voids can be underestimated 
by up to 20\%. 

We confirm the previous findings by 
\citet{Grogin00, Hogg04, Rojas05, Croton05, Patiri06b}, that in 
lower-density regions, the galaxy population is dominated by blue actively
star forming galaxies. We fitted the $B_J-R$ color distribution of 
2dFGRS void galaxies by double Gaussian distributions. There are significantly 
more void galaxies in the blue cloud than in the general field, and the 
red sequence is strongly suppressed. In addition, we have found the 
indication that 
the blue population of galaxies in void centers is slightly {\it bluer} 
than the field population, not just more numerous. The 
shift is only minor and almost of the same order as the maximal 
quoted uncertainty by \cite{Cole05}, so it remains unclear how robust is this 
effect. It is not reproduced by the semi-analytical models for galaxy 
formation of the Millenium simulation. Also the red sequence in the 
Millenium catalogue has a significantly tighter spread. This may due to 
a too sharp cutoff of the star formation activity in the semianalytical 
models which seems to be unrealistic. 

The radial distribution of the ratio of early and late type galaxies
inside voids coincides quantitatively for the 2dFGRS and the Millenium 
catalogue, but in the data, the fraction of E-galaxies stays smaller 
(and the fraction of S-galaxies larger) further outwards beyond the 
boundary of the voids. The abundance of star-forming galaxies is higher 
in the 2dFGRS voids than in the semianalytical model. 

\section*{Acknowledgments}

We thank the 2dFGRS team for the excellent data they provided. Part of this
project was done at the Aspen Center for Physics workshop on cosmic voids.
We thank the participants for stimulating discussions, especially Daren 
Croton concerning the Millenium simulation, and Michael Vogeley,  
Rien van de Weygaert, Ravi Sheth and Jeremy Tinker on the void statistics. 
Advice from Paco Prado and Santiago Patiri are gratefully acknowledged. 
We thank the anonymous referee for his constructive comments. 

\bibliography{Biblio/articles}
\bsp

\label{lastpage}

\end{document}